\documentclass[graybox]{svmult}

\usepackage{geometry}
\usepackage{ebgaramond,ebgaramond-maths,makeidx,graphicx,multicol}
\usepackage[bottom]{footmisc}
\makeindex

\begin{document}

\title*{Temporal Networks as a Modeling Framework}

\author{Petter Holme and Jari Saram\"aki}

\institute{Petter Holme \at IceLab, Department of Physics, Ume{\aa} University, 90187 Ume{\aa}, Sweden.\\ \email{petter.holme@physics.umu.se} \newline  Department of Energy Science, Sungkyunkwan University, Suwon 440--746, Korea \and Jari Saram\"aki \at Department of Biomedical Engineering and Computational Science, School of Science, Aalto University, 00076 Aalto, Espoo, Finland.\\ \email{jari.saramaki@aalto.fi}}

\maketitle

\abstract*{To understand large, connected systems we cannot only zoom into the details. We also need to see the large-scale features from afar. One way to take a step back and get the whole picture is to model the systems as a network. However, many systems are not static, but consisting of contacts that are off and on as time progresses. This Chapter is an introduction to the mathematical and computational modeling of such systems, and thus an introduction to the rest of the book. We will cover some of the earlier developments that form the foundation for the more specialized topics of the other Chapters.}

\abstract{To understand large, connected systems we cannot only zoom into the details. We also need to see the large-scale features from afar. One way to take a step back and get the whole picture is to model the systems as a network. However, many systems are not static, but consisting of contacts that are off and on as time progresses. This Chapter is an introduction to the mathematical and computational modeling of such systems, and thus an introduction to the rest of the book. We will cover some of the earlier developments that form the foundation for the more specialized topics of the other Chapters.}

\section{Introduction}
\label{sec:intro}
Life, at many levels, is about large connected systems. In the biological sense, life is a consequence of macromolecules building cells and carrying information. More mundanely, our everyday life happens in amid a network of friends, acquaintances and colleagues. To understand life, at every level, we need to zoom out from macromolecules or friendships and look at their global organization from a distance. Here, zooming out means discarding the less relevant information in a systematic way. One approach to this, successful the last decade, is network modeling. This means that one focusses on the units of the system, be it proteins or persons, and how they are connected, and nothing else. Of course, this is a very strong simplification. One often has more information about a system that would enrich rather than obscure the picture. One such additional type of information regards when the interactions happen between the units. The essence of temporal-network modeling is to zoom out by excluding all information except which pairs of units that are in contact and when these contacts happen.

There is a large number of systems that could, potentially, be modeled as temporal networks. In addition to the cellular processes and social communication mentioned above large technological infrastructures---technologies based on the Internet or mobile-phone networks for example---have both network and time aspects that make them interesting for temporal network modeling. Neural networks---perhaps primarily at a functional level of brain regions that are considered connected if there is a measurable information transfer between them---are another example. A third example is ecological networks and species and their interaction. Such networks---like food webs, depicting which species feed on which other species, or mutualistic networks of species  providing mutual benefits, such as plants and pollinators---experience time varying changes with the seasons and other environmental changes.

\begin{figure}[b]
\sidecaption
\includegraphics[width=0.5\linewidth]{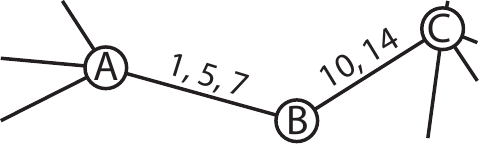}
\caption{Illustrating the non-transitive nature of temporal networks. Information can spread from A to B, from B to C, from C to B, from B to A, it can also spread from A to C via B, but not from C to A since by the time the information can have reached B all contacts between A and B have already happened. Note that static networks are transitive, even directed networks, so one cannot simply reduce a temporal network to a static one.}
\label{fig:non_trans}
\end{figure}

In this chapter, we will review the essential mathematical and computational techniques for extracting information from a temporal network representation of a system. We will discuss quantitative measures of network structure, computational techniques of successive randomization to study these measures, and models to generate and explain temporal networks and studies seeking to explain the effects of the temporal-topological structures on dynamics taking place on the networks. For a more comprehensive review of the field, see Holme and Saram\"aki~\cite{holme_saramaki}.

\section{Measuring temporal network structure}
\label{sec:measuring}

In this section, we will review some of the proposed structural measures that strive to capture both temporal and topological features and correlations. For the rest of the chapter, we will consider systems that can be represented as lists of contacts---triplets of pairs of vertices together with the time of their contact, or alternatively as quadruples containing the beginning and end times of contacts, if these are not instantaneous. We call the first type of temporal network a  \emph{contact sequence}, the other one an \emph{interval graph}.

\begin{figure}[b]
\sidecaption
\includegraphics[width=0.5\linewidth]{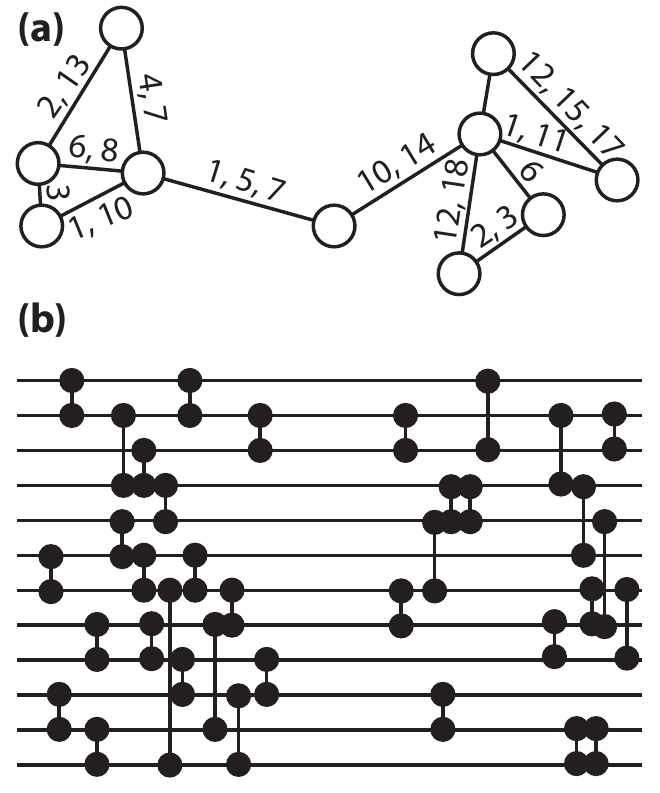}
\caption{Visualization of temporal networks. (a) shows a labeled aggregate network where the labels denote times of contact, and (b) shows a time-line plot, where each of the lines corresponds to one vertex and time runs from left to right.}
\label{fig:visualization}
\end{figure}

Before we start discussing effects of structural measures, we note that temporal networks are notoriously difficult to visualize in a way that would both show all possible temporal information and highlight the important structural features (similarly to what e.g.\ spring-embedding can successfully do for static networks). Two representations, labeled graphs and time-line plots, are illustrated in Fig.~\ref{fig:visualization}. Of these, the time-line plots can help to visualize the temporal structure (including temporal heterogeneities such as bursts)\ while the labeled graphs highlight the network topology. However, neither of them can be scaled up to more than a dozen or so vertices. There are other attempts of combining time and simplified topology---most notably the \emph{alluvial diagrams} of Rosvall and Bergstrom~\cite{rosvall_bergstrom}---that however typically miss the non-transitivity of temporal networks, or some other important aspects.

\subsection{Reachability and latency}

\begin{figure}[b]
\sidecaption  
\includegraphics[width=0.5\linewidth]{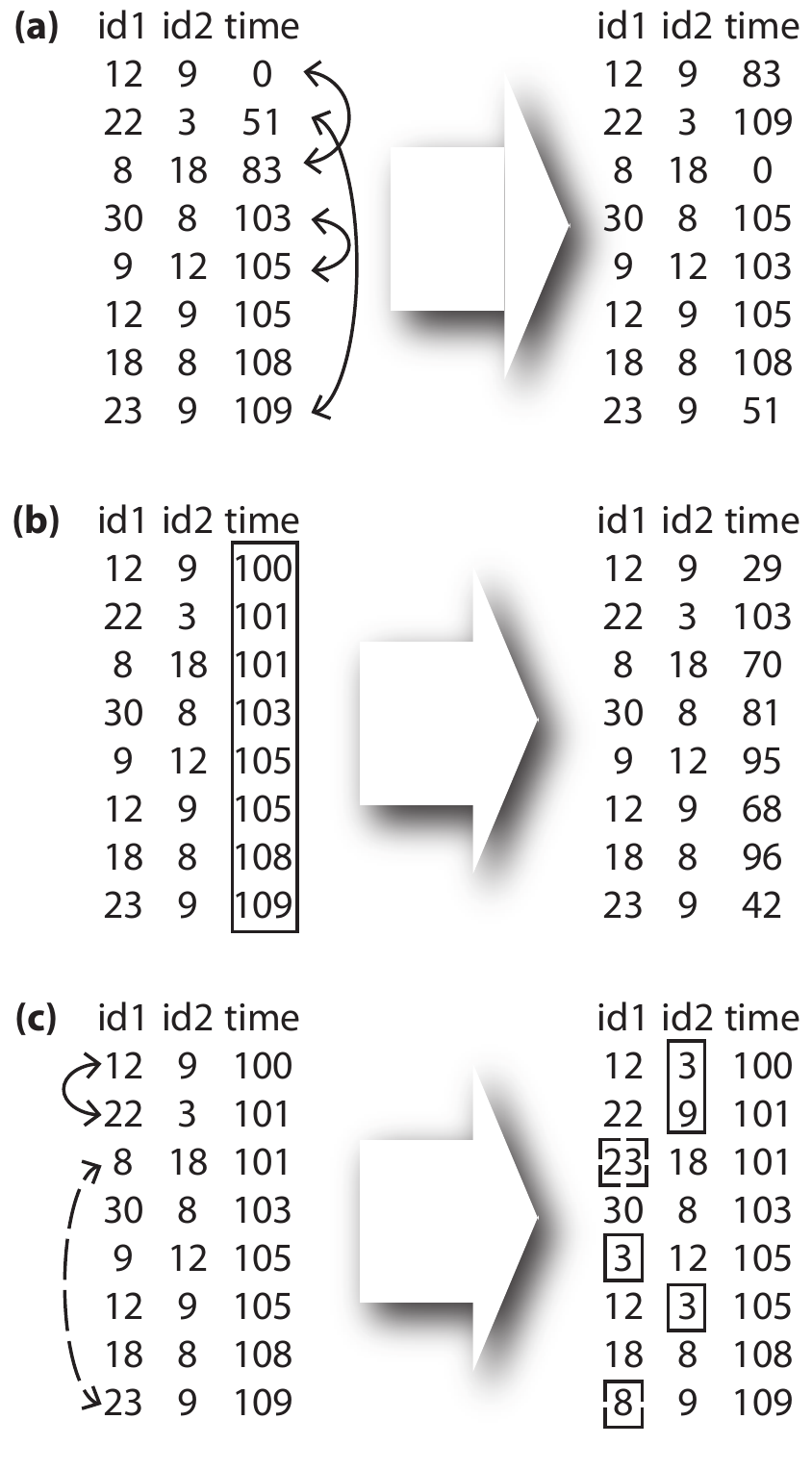}
\caption{Illustration of some randomization methods. (a) shows the randomly permuted times (PT) scheme that removes structures in the order of events. (b) shows the random times scheme (RT) and (c) displays a static network rewiring as it appears in a contact sequence. The contacts of an edge is conserved in the process. Note that one need to allow edges (A,B) and (C,D) to be rewired to both (A,D), (C,B) and (A,C), (D,B) to make the sampling ergodic.}
\label{fig:intrans}
\end{figure}

One of the most fundamental differences between temporal and static networks is that the former are not transitive. Even if vertex A is related to B and B is related to C, it might be that A is not related to C (see Fig.~\ref{fig:intrans}). The relation in question is in essence the possibility of something spreading from one vertex to another through a series of contacts where the times of the contacts are increasing (anything else would not be feasible in reality). For this reason, the statistics of such \emph{time-respecting paths} are very informative. Authors have e.g.\ investigated the average durations of time-respecting paths~\cite{holme_2005,kossinets_etal,tang_etal_2010,pan_saramaki}. Given a pair of vertices $(i,j)$ and a time $t$, the \emph{latency} is the shortest time to go from $i$ to $j$ at time $t$ following only time-respecting paths.

The latency is not the entire story, since just like regular graphs, temporal networks can be disconnected. This is, in practice, more common in temporal than static networks, as the paths joining vertices need to be traversed in the order of contacts. 
A practical measure for capturing this would be the expected value of the number of vertex pairs that have infinite latency values~\cite{holme_2005}.
For empirical data, the finite period of observation may also play a role, because paths whose realization takes a very long time may not ever be completed within the observed period. It should be emphasized that connectivity is only defined within some time window: the fact that A is connected to B via a time-respecting path that begins at $t$ does not guarantee that such a connection exists at any later point in time.  Furthermore, connectivity is always directed as dictated by the arrow of time (see the transitivity example above).

One can elaborate on latency-like measures in many ways in order to capture different aspects of  reachability and dynamical influence between nodes. It could for example be interesting to monitor the number of time-respecting paths between pairs of vertices in order to capture frequently appearing pathways, or to resolve the average latency in time---it might be that the average latency follows e.g.\ a daily pattern where time-respecting paths are faster to traverse during the time of  day when the contacts are more frequent. Additional constraints may also be set on time-respecting paths; e.g.\ one may require that the contacts forming a path follow each other rapidly enough, so that long waiting times between contacts destroy the path~\cite{pan_saramaki}.

\subsection{Clustering and correlations}
In static networks, the local structure---focusing on the immediate surroundings of an average vertex---is an important predictor of the behavior of dynamic systems on the network. Adding the temporal dimension is not straightforward, which perhaps explains the rather few attempts to do so. Below, we sketch one possible approach and illustrate some of the inherent difficulties.

In static networks, the level of connectivity in the neighborhood of a vertex can be measured with the (local) topological clustering coefficient.  Its values are normalized, such that a value of 0 indicates no connectivity and a value of 1 the existence of all possible connections. Adding a temporal dimension, we would like to measure the connectedness of a neighborhood around a given moment of time $t$. In other words, we would like to put a larger weight on contacts that are closer to $t$ in time. This can be done by an summation kernel $F(t)$ with the properties that it is bounded, non-negative,  monotonically increasing for $t < 0$, monotonically decreasing for $t > 0$, and $F(0) = 1$. A temporal clustering coefficient for a contact sequence would then be the following sum divided by some normalizing factor
\begin{equation}\label{eq:normalizing}
\sum_{c,c',c''}F(t(c)-t)\,F(t(c')-t)\,F(t(c'')-t)
\end{equation}
where $c=(i,j,t)$, $c'=(j,k,t')$ and $c''=(k,i,t'')$ are contacts with $i\neq j\neq k$.  However, there is no obvious choice of denominator to balance this with. If one calculates the denominator by assuming that there can be one contact per time step between all vertices that are in contact with $i$ at some point, then this would for the very most datasets be a number many orders of magnitude larger than the denominator. One can perhaps set the maximal number of contacts as the total number of contacts in the dataset and assume that  they all happened at time $t$, but also this would in practice be a very much larger number than that given by Eq.~\ref{eq:normalizing}. A third option would be to replace the $F$-factors by $1$ (their maximal value), but this would be equal to assuming that the number of contacts per vertex pair that has been in contact at some time is fixed, which would be strange for most types of temporal networks. This example is not meant to discourage from constructing measures capturing both temporal and topological structures, but rather the other way around. It shows how moving away from the assumption that all edges are equivalent (an assumption underlying most static network representations) requires new ways to think about network concepts. In this case, the best solution, we believe, would be to compare the raw sum to that obtained from a carefully chosen reference model.

In static networks, one important class of measures quantifies the relationship between the degrees of connected nodes. Is there an overrepresentation of edges between, say mid-degree vertices and other mid-degree vertices? Such an analysis can be made at different levels, from plotting the entire correlation profile~\cite{maslov_sneppen} to measuring a scalar-valued correlation coefficient~\cite{newman_2010}. These degree-correlation measures can be generalized to temporal networks more straightforwardly  than the clustering coefficient. One can use similar summation kernels as discussed above to replace node degrees by a time-dependent activity level, and then perform the same analysis. Then again, while this would capture something similar in spirit to the degree-correlation measures designed for static networks, in temporal networks there is a multitude of other conceivable concepts of correlations across links that could well prove more important.

As a temporal networks  evolve, some subsets of their nodes and links may be more continuously active than others. Such persistent patterns are subnetworks that are prime candidates for functional subunits of some sort; they could also be an interesting alternative to aggregating all contacts if one wants to reduce the system to a static network. As an example of how to investigate persistent contact patterns, one can let a time window slide through an interval graph and calculate the adjacency \emph{correlation function}, or \emph{vertex persistency}
\begin{equation}\label{eq:persistency}
\gamma_i(y)=\frac{\sum_{j\in\phi(i,t)}a(i,j,t)a(i,j,t+1)}{\sqrt{\sum_{j\in\phi(i,t)}}\sqrt{\sum_{j\in\phi(i,t+1)}}}
\end{equation}
where $t$ is the beginning of the time window, and $\phi(i,t)$ are the non-zero indices of the (time-dependent) adjacency matrix.

\subsection{Centrality}

Network centrality measures form one of the most important classes of measures of static network structure. These quantities try to capture various facets of the question how central a vertex is.  Ref.~\cite{pan_saramaki} discusses a centrality measure akin to \emph{closeness} in static graphs~\cite{newman_2010}. Essentially, the authors define centrality as the average reciprocal value of the time from the focal vertex to all others. Tang \textit{et al.}\ also defines a (somewhat different) closeness centrality for temporal networks together with a temporal version of the \textit{betweenness centrality}~\cite{tang_etal_2010b}.

Takaguchi \textit{et al.}\ take a slightly different approach in Ref.~\cite{takaguchi_etal2} when they measure a kind of influence (called \emph{advance}) related to concepts of centrality by focusing on the importance of the events the vertex participates in. This work draws on previous results from Ref.~\cite{kossinets_etal}. Mantzaris \textit{et al.}~\cite{mantzaris_etal} use a spectral centrality measure for temporal graphs to study learning in the human brain.

\subsection{Motifs}
Network motifs were first proposed for static directed networks~\cite{alon} and are, briefly described, overrepresented subgraphs formed by a few vertices and their directed links. Motifs are often interpreted as functional units, or candidates for such, and motif analysis is commonly applied e.g.\ in systems biology. In static directed networks, motifs can be mapped to component-like structures such as feedforward loops, but in temporal networks, this is harder. Rather, motifs in temporal networks correspond to typical sequences of events. There are many ways of defining such motifs. To take one example, Kovanen \textit{et al.}~\cite{kovanen2011} look at  sequences of contacts between vertices that are maximally separated by a time $\delta t$. More precisely, two contacts $e_i$ and $e_j$ are $\delta t$-adjacent if they share a vertex and are separated in time by no more than $\delta t$. Pairs of events are then defined as  $\delta t$-connected if there is a sequence of $\delta t$-adjacent events joining them and temporal subgraphs are defined as sets of events that are $\delta t$-connected. By counting such subgraphs and mapping them into isomorphic classes on the basis of their order of events, Kovanen \textit{et al.} find an overrepresentation of temporal motifs that are causal, i.e.\ where the contacts may have triggered one another (such as A contacts B who contacts C and D, as opposed to the non-causal sequence where B contacts C and D, and A only then contacts B). As an application of a temporal-network motif method (slightly different from that of Ref.~\cite{kovanen2011}) Jurgens and Lu~\cite{jurgens_lu} use motifs to study behavior in the evolution of Wikipedia.

\subsection{Mesoscopic structures}
In static networks, there has recently been a flood of methods proposed to discover mesoscopic structures (a.k.a. clusters, communities or modules~\cite{fortunato}). These are loosely defined as groups of vertices more densely connect within than between each other. Much of the community structure literature regarding static networks only focuses on deriving a method for decomposing the network on the basis of some conceptually simple principle. The few methods that incorporate the time dimension into  community detection typically operate on aggregated time-slices of the contact sequence~\cite{braha,braha_bar_yam} or networks of links that have happened and will happened again~\cite{holme_2003}. One can imagine clustering algorithms based on more elaborate temporal structures, like time-respecting paths (an exception is Lin \textit{et al.}~\cite{lin_etal}). As mentioned earlier, visualizing temporal networks as two-dimensional, printable diagrams is difficult and this is a major obstacle to intuitive reasoning about mesoscopic temporal-topological structure. Reducing the network to a network of clusters that split and merge with time is perhaps the most promising path in this direction. Unfortunately, such a reduction would also destroy any non-transitive features of the original structure, especially when time slices or aggregation are involved, and smear out the effects of all temporal structures associated with  shorter time scales than the time window that is used (such as bursts).~\cite{krings_etal}.

\section{Models of temporal networks}
As in all other areas of theoretical science, our understanding of temporal networks hinges on mathematical and computational models. These models have different purposes. The simplest class of models, already mentioned above, involves \emph{null} or \emph{reference models} that are used together with various measures in order to infer their statistical significance, or in order to assess the contribution of chosen types of correlations to the values of the measures. Related to this are generative models that can serve both as reference models and as a method to synthesize temporal structures to run simulations of dynamic systems on. The third class comprises mechanistic models for explaining the emergent network structures that one measures; and finally we also have predictive models that are tailored to forecast future aspects of a temporal network.

Much remains to be done in the development of temporal-network models in all  above-mentioned areas. This is somewhat in contrast to the theory of static networks where a very large number of models were developed at an early stage~\cite{newman_2010}.

\subsection{Randomized null or reference models}
In order to interpret the significance of temporal-network measures, or to understand what effects different temporal and structural features have on them,  one needs something for comparing against. One approach is to compare observations against the same measures computed for an ensemble of randomized networks that are "neutral" in some chosen sense, i.e.\ where certain correlations and features are removed by randomization procedures such that only the chosen fundamental constraints of empirical networks are retained. 
The way the empirical network differs from the null model gives meaning to the structure observed in the empirical data, and points out effects that are due to the removed correlations. One can also give a scale to the "raw" measures of temporal network structure, e.g.\ by normalizing by the corresponding values for the reference models or subtracting such values. More elaborately, one can assess the statistical significance of observed measures by calculating the $Z$-score, or similar, of their values in the empirical network against their distribution in the randomized ensemble.
Strictly speaking, the term "null model" should be used only in this context:  $Z$-scores are then used to reject or verify the null hypothesis that the observed features can be explained by the randomized ensemble, and the removed correlations do not play a role.

For static networks, perhaps the most popular reference model is to rewire the edges randomly while keeping the degree and number of vertices constant~\cite{maslov_sneppen}. See Fig. \#. This model is very closely related to the \emph{the configuration model} which is rigorously defined as an ensemble of vertices of given degrees that are connected in a maximally random way~\cite{newman_2010}. 
For practical purposes,  randomization by edge swapping and the configuration model are equivalent. This is helpful since the configuration model is theoretically well understood.

For temporal networks, randomization plays a slightly different role than for static networks. On one hand, there is no mathematically well-understood null model like the configuration model. On the other hand, there is a much larger number degrees of freedom, and subsequently there is also a much larger number of possible randomization models.  In fact, such models allow for going through a sequence of randomization procedures, where one progressively removes chosen types of correlations from the empirical temporal network. This provides a detailed, continuous picture of temporal-network structure and the correlations that underlie observations.

 Below, we will discuss some of the most important temporal network models. For a longer list, see Holme and Saram\"aki~\cite{holme_saramaki}.

\subsubsection{Randomly permuted times (RP)}
As a temporal counterpart to the randomization of edges discussed above, one can randomly permute the times of contacts, while keeping the network's static topology and the numbers of contacts between all pairs of vertices fixed. As this randomization scheme retains all static network structure and the number of contacts for each edge, its application can be used to study the effects of the exact order and timings of contacts. The reference model destroys burstiness and inter-contact times of nodes and edges, and subsequently also all correlations between timings of contacts on adjacent edges. The model also keeps the overall rate of events in the network over time, such as daily or weekly patterns.

\subsubsection{Random times (RT)}
The set of time stamps is conserved in the RP ensemble. Hence, the intensities of contacts at an aggregate level follow the same patterns as the real data---if there are circadian rhythms (like in many human, and other biological and social systems) they will still be there in randomized networks. If one wants to explore the impact of these rhythms, one may draw the interaction times from a random distribution and compare the outcome to the RP ensemble. 

\subsection{Generative, mechanistic and predictive models}
A large amount of numerical temporal network studies have been performed on empirical networks. Since there is not yet any commonly-agreed-upon set of temporal network characteristics, and since it is not yet clear what the most important features of temporal networks are and whether there is any universality in such features, the focus has been on what the data can tell us, which can be seen as an advantage. The disadvantage is that we do not get a systematic understanding of the effects of temporal network structure. In order to arrive at such an understanding, we need generative models that can output temporal networks with tunable structure.

In cases where topological and temporal structures are decoupled, creating a generative model is straightforward. One could perhaps first generate the topology according to some model from the static network literature, and then generate time series of contacts over the edges for tuning some quantity of interest (like the vertex persistency, Eq.~\ref{eq:persistency}). Studies that successfully utilize generative models to study the effects of temporal network structure include Refs.~\cite{perra_etal} and \cite{rocha_etal_2012}.

Another type of models that are very common in the static network literature but not for temporal networks are mechanistic models trying to explain the emergence of large-scale structure from simple underlying driving mechanisms. Indeed, the whole field of complex network theory took off in the 1990's with the Watts--Strogatz model of small-world networks and the Barab\'asi--Albert model of scale-free networks, which makes this lack of mechanistic temporal network models even more conspicuous.

A third type of models  for temporal networks  (largely still waiting to be realized) is predictive models, solely targeted at forecasting the future development of the contact structure. Such models, drawing from machine-learning and statistical techniques, would not necessarily attempt to explain why a temporal network is like it is, or to generate contact sequences from scratch. Rather, given a contact sequence or interval graph, such models could  predict its continuation in the near future.

\section{Processes on temporal networks}
Networks are never just a collection of vertices and edges (or contacts in the case of temporal networks), except in very trivial cases. Rather, they are the underlying structure that determines how dynamical processes over the graph unfold, from contagion of infections to Internet traffic. Thus, they in reality define the system's function. Obviously, temporal effects can strongly affect any dynamics that follow the shortest paths between vertices (see the above discussion on latency). Especially, temporal features of networks affect the dynamics of diffusion and spreading. This has been investigated by comparing spreading dynamics---often, with the help of compartmental models of infectious disease spreading---on empirical contact sequences and their randomized reference counterparts. At the moment, we do not have a comprehensive theory of how temporal-network structure affects disease spreading. For some systems the temporal structure speeds up spreading~\cite{karsai_etal}, in other systems the temporal structure seems to slow it down~\cite{rocha_etal_2011}. The structure in focus of these studies is \emph{burstiness}---the property that contact activity (often human) is very inhomogeneously distributed in time---that can be readily removed from temporal networks by applying randomized reference models. 

Another type of models of social spreading phenomena is threshold models, in particular targeted for studies of social influence and opinion spreading. In threshold models, an agent (or vertex) changes its state whenever the impact from the surrounding vertices exceeds some threshold value. The dynamics of such models seem to have a tendency to speed up if there is  burstiness, but also in this case there is no general theory. Coming up with a general theory might even be impossible, since  there are different aspects of how to measure and quantify the impact that triggers state changes in such models. Furthermore, it is hard to observe or experimentally study human threshold behavior, which additionally may largely depend on circumstances (for one such experiment, see~\cite{CentolaScience2010}). Regarding threshold models on temporal networks, Karimi and Holme~\cite{karimi_holme} studied a modification of Watts's threshold models of cascades~\cite{watts} for contact sequences and Takaguchi, Masuda and Holme~\cite{takaguchi_etal} studied a threshold model of exponentially decaying influence. Both these studies were performed on empirical networks.

\section{Summary and discussion}
We have given an overview of the different aspects that the field of temporal network so far has covered. Furthermore, we have explained the challenges in extending static network measures to temporal networks. We argue these challenges should be encouraging rather than the opposite, both since they are intellectually fascinating and since there are useful applications waiting once they are resolved.

The study of temporal networks is a fast advancing field with a great potential for the future. However, many challenges remain. The extra level of information added by the temporal dimension does not only make it difficult to develop theory and computational methods, it also changes the questions one can ask about the structure of the system. Probably many advances can be made by connecting and integrating temporal networks with other extensions of network models such as \emph{spatial networks}~\cite{barthelemy} where the coordinated of spatially embedded nodes and links are incorporated in the modeling framework, or \emph{adaptive networks}~\cite{sayama_gross,braha_bar_yam} where there is a focus on the feedback from the dynamics on the networks to the evolution of the network evolution.

Open questions for future studies ranges from how to make static visualizations of temporal networks, via how to predict missing links in incomplete temporal network data~\cite{sarkar_etal} or how to make sport-ranking systems~\cite{motegi_masuda}, to classic questions like if there is any universal law that involves both temporal structure and network topology.

\section{In this book}

This book aims at presenting an overview of the state-of-the-art in temporal networks. Its chapters are contributed by leading researchers and research teams from a variety of backgrounds and disciplines. Our target has been to cover the emerging field of temporal networks both in breadth and in depth, and because of this, some chapters are essentially reviews on key topics---such as  temporal network metrics and burstiness---whereas others provide detailed accounts of investigations building on the temporal networks framework, from networks of face-to-face human proximity to the collective behaviour of social insects.

The following five chapters focus on metrics, measures and methods for characterizing temporal network structure. In Chapter 2, Nicosia, Tang, Mascolo, Musolesi, Russo, and Latora present an overview of  key temporal network metrics and measures, from walks and paths to connectedness and centrality measures. This is followed by a chapter that focuses on one of the key characteristics of temporal networks, especially those related to human interactions: in Chapter 3, Min, Goh, and Kim discuss burstiness, from measuring and characterizing bursty activity to modeling its origins and assessing its effects on dynamical processes. Then, in Chapter 4, Caceres and Berger-Wolf address the important problem of the underlying temporal scales in interaction streams that define temporal networks, focusing on identifying inherent temporal scales and finding network representations that match those scales. These overviews are followed by two chapters that focus on specific temporal network features and measures: Zhao, Karsai and Bianconi discuss the entropy of temporal networks in Chapter 5,  combining modeling efforts with studies of large, time-stamped empirical data sets. In Chapter 6, Kovanen, Karsai, Kaski, Kert\'{e}sz and Saram\"aki present the temporal motifs approach that is designed to detect, categorize and quantify recurrent temporal mesoscopic patterns of link activations.

In the second part of the book, temporal network metrics and measures are put to use in empirical studies. In Chapter 7, Tang, Leontiadis, Scellato, Nicosia, Mascolo, Musolesi, and Latora apply the metrics discussed in Chapter 2 to the analysis of a number of empirical and simulated data sets. Then, in Chapter 8, Vazquez returns to the topic of burstiness, and analyzes how bursty dynamics impacts spreading processes in computer and social systems. The effects of burstiness on spreading processes are further studied in Chapter 9 by Miritello, Lara, and Moro
in the context of networks of human interactions, and connected to the social, topological structure around individuals. This is followed by an account of temporal networks of face-to-face human interactions by Cattuto and Barrat in Chapter 10; spreading dynamics are also used here to probe the temporal structure of proximity patterns. Finally, in Chapter 11, Charbonneau, Blonder, and Dornhaus present an inspiring overview of social insects as model systems for network dynamics, and discuss how temporal network analysis methods can provide novel ways to view the complexity of collective behavior of social insects.

The third and last part of the book discusses models of temporal networks and  processes taking place on such networks. In Chapter 12, Masuda, Takaguchi, Sato, and Yano consider the origins of the long-tailed inter-event interval distributions in human dynamics, and model them with the Hawkes process, a self-exciting point process that is fitted to data on face-to-face interactions in company offices. Mantzaris and Higham then address the micro-scale dynamics of triadic closure in social networks with the help of a model and time-stamped electronic records in Chapter 13. The same authors then move on to dynamic communicability measures, and show that they can be used to predict macro scale features of simulated epidemics on temporal networks in Chapter 14. The last three chapters focus on the behavior of other archetypal dynamic processes than spreading, when the dynamics unfolds through the interactions sequences of temporal networks. In Chapter 15, Hoffmann, Porter and Lambiotte develop a mathematical framework for random walks on temporal networks using an approach that provides a compromise between abstract but unrealistic models and data-driven but non-mathematical approaches. Karimi and Holme then develop and study a version of Watts's cascade model for the spreading of  opinions and innovations in the temporal network setting in Chapter 16, and finally, Fern\'andez-Gracia, Egu\'iluz, and San Miguel present version of the Voter model of opinion dynamics that is able to account for heterogeneous temporal activity patterns in Chapter 17.

\begin{acknowledgement}
The authors acknowledge financial support by the Swedish Research Council (PH), the WCU
program through NRF Korea funded by MEST R31--2008--10029 (PH), EU's 7th Framework
Program's FET-Open to ICTeCollective project no.\ 238597 (JS) and the Academy of Finland,
the Finnish Center of Excellence program 2006--2011, project no.\ 129670 (JS).
\end{acknowledgement}

\bibliographystyle{abbrv}
\bibliography{holme_saramaki}

\end{document}